\def\deg{\ensuremath{^\circ}}

\documentclass{aa}
\usepackage{epsfig}
\begin{document}

\title{Early SPI/INTEGRAL measurements of 511 keV line emission from
the 4$^{\rm th}$ quadrant of the Galaxy\thanks{Based on observations with
INTEGRAL, an ESA project with instruments and science data centre
funded by ESA member states (especially the PI countries: Denmark,
France, Germany, Italy, Switzerland, Spain), Czech Republic and
Poland, and with the participation of Russia and the USA.}}

\author{P.~Jean\inst{1}
         \and J.~Kn\"odlseder\inst{1}
         \and V.~Lonjou\inst{1}
         \and M. Allain\inst{1}
         \and J.-P.~Roques\inst{1}
         \and G.K.~Skinner\inst{1}
         \and B.J.~Teegarden\inst{1,2}
         \and G.~Vedrenne\inst{1}
         \and P.~von~Ballmoos\inst{1}
	 \and B.~Cordier\inst{6}
	 \and R.~Diehl\inst{3}
	 \and Ph.~Durouchoux\inst{6}
         \and P.~Mandrou\inst{1}
         \and J.~Matteson\inst{7}
         \and N.~Gehrels\inst{2}
         \and V.~Sch\"onfelder\inst{3}
	 \and A.W.~Strong\inst{3}
	 \and P.~Ubertini\inst{5}
	 \and G.~Weidenspointner\inst{1,2}
	 \and C.~Winkler\inst{4}
}

\offprints{Pierre Jean, e-mail : jean@cesr.fr}

\institute{
$^{1}$ CESR, CNRS/UPS, B.P.~4346, 31028 Toulouse Cedex 4, France \\
$^{2}$ LHEA, NASA/Goddard Space Flight Center, Greenbelt, MD 20771, USA \\
$^{3}$ MPI f\"ur Extraterrestrische Physik, Postfach 1603, 85740 Garching, Germany  \\
$^{4}$ ESA-ESTEC, RSSD, Keplerlaan 1, 2201 AZ Noordwijk, The Netherlands \\
$^{5}$ IAS, CNR, Via del Fosso de Cavaliere 00133 Roma, Italy \\
$^{6}$ DSM/DAPNIA/SAp, CEA-Saclay, 91191 Gif-sur-Yvette, France \\
$^{7}$ CASS, UCSD, La Jolla, CA 92093, USA 
}

\date{Received / Accepted }

\authorrunning{P.~Jean et al.}

\titlerunning{SPI/INTEGRAL measurements of galactic e$^{+}$e$^{-}$ emission}

\abstract{ We report the first measurements of the 511~keV line
emission from the Galactic Centre (GC) region performed with the
spectrometer SPI on the space observatory INTEGRAL
(International Gamma-Ray Astrophysics Laboratory). Taking into account the
range of spatial distribution models which are consistent with the
data, we derive a flux of $9.9^{+4.7}_{-2.1} \times 10^{-4}$ ph
cm$^{-2}$ s$^{-1}$ and an intrinsic line width of
$2.95^{+0.45}_{-0.51}$ keV (FWHM). The results are consistent with
other high-spectroscopy measurements, though the width is found to
be at the upper bound of previously reported values.
\keywords{Gamma rays: observations -- Line: profiles -- Galaxy: center} }

\maketitle

\section{Introduction}
\label{sec:intro}

Line emission at 511 keV from the GC region has been observed
since the early seventies in balloon and satellite experiments. The
line was discovered at an energy of 476 $\pm$ 26 keV (Johnson et~al.
1972), so the physical process behind the emission was initially ambiguous
and firm identification had to await the advent of high resolution
spectrometers.  In 1977, germanium
(Ge) detectors, flown for the first time on balloons, allowed the
identification of the narrow annihilation line at 511~keV.
Its width turned out to be only a few keV (Albernhe et~al. 1981;
Leventhal et~al. 1978). The eighties were marked by ups and downs in
the measured 511 keV flux through a series of observations performed by the
balloon-borne Ge detectors (principally the telescopes of
Bell-Sandia and GSFC). The fluctuating results were interpreted as the
signature of a compact source of annihilation radiation at the GC (see e.g. Leventhal 1991).  
Additional evidence for this scenario
came initially from HEAO-3 (Riegler et~al. 1981) reporting
variability in the period between fall 1979 and spring 1980. Yet, during
the early nineties, this interpretation was more and more questioned, since
neither eight years of SMM data (Share et~al. 1990) nor the revisited
data of the HEAO-3 Ge detectors (Mahoney et~al. 1993) showed evidence
for variability in the 511~keV flux.  Throughout the nineties, CGRO's
Oriented Scintillation Spectrometer Experiment (OSSE) measured steady
fluxes from the galactic bulge and disk component and some spatial 
information about annihilation emission was extracted from 
the OSSE, SMM and TGRS experiments (Purcell et~al. 1997; Milne 
et al. 2001). A possible
additional component
at positive galactic latitude was attributed to an annihilation
fountain in the GC (Dermer \& Skibo 1997).  A summary of
the key parameters measured by high resolution spectrometers is given in 
Table 1.

\begin{table*}
\label{historic measurements}
     \caption{The GC 511 keV line measured by more recent high 
resolution
spectrometers. The uncertainties quoted for the SPI results include
the effects of the range of models discussed in the text (section 3).}
     \begin{array}[b]{lccccl}
\noalign{\smallskip}
\hline
\hline
\noalign{\smallskip}
\mbox{instrument}  & \mbox{year} & \mbox{flux} & \mbox{centroid} & 
\mbox{width (FWHM)} & \, \mbox{  references}\\
             &      & [10^{-3}\ ph\ cm^{-2}\ s^{-1}]  & [keV] & [keV] \\
\hline
\noalign{\smallskip}

\mbox{HEAO-3$^{a}$}   & 1979-1980   & 1.13 \pm 0.13 & 510.92 \pm 0.23 & 1.6^{+0.9}_{-1.6}   & \mbox{Mahoney et~al. 1994}\\
\mbox{GRIS$^{b}$}     & 1988 \ and \ 1992  & 0.88 \pm 0.07 &          & 2.5 \pm 0.4   & \mbox{Leventhal et~al. 1993}\\
\mbox{HEXAGONE$^{b}$} & 1989        & 1.00 \pm 0.24 & 511.33 \pm 0.41 & 2.90^{+1.10}_{-1.01} & \mbox{Smith et~al. 1993}\\
\mbox{TGRS$^{c}$}     & 1995-1997   & 1.07 \pm 0.05 & 510.98 \pm 0.10 & 1.81 \pm 0.54 & \mbox{Harris et~al. 1998}\\
\mbox{SPI}     & 2003   & 0.99^{+0.47}_{-0.21}  & 511.06^{+0.17}_{-0.19} & 2.95^{+0.45}_{-0.51} & \mbox{this work}\\

\noalign{\smallskip}
  \hline
\end{array}
\mbox{$^{a}$ assuming a point source.}

\mbox{$^{b}$ flux in the field of view (17\deg\ and 18\deg\ for GRIS 
and HEXAGONE respectively).}

\mbox{$^{c}$ gaussian-shape source (FWHM 30\deg).}

\end{table*}

We report the first measurements of galactic 511~keV
gamma-ray line emission performed by the spectrometer SPI on INTEGRAL.
SPI is a coded-mask telescope with a fully-coded field of view of
16\deg\ that uses a 19 pixel high-resolution Ge detector 
array resulting in an
effective area of about 75~cm$^2$ at 511~keV (Vedrenne et~al. 1998).

\section{Data analysis}
\label{sec:analysis}

The data analysed in this work were accumulated during the first
year's GC Deep Exposure (GCDE) and Galactic Plane
Scan (GPS), executed as part of INTEGRAL's guaranteed time
observations (see Winkler 2001). We used data from 19 orbits from
March 3rd to April 30th, 2003, amounting to a total effective
exposure time of 1667 ks. The GCDE consists of rectangular
pointing grids covering galactic longitudes $l=\pm 30^\circ$ and
latitudes $b=\pm 10\deg$, with reduced exposure up to
$b=\pm 20\deg$. The GPS consists of pointings within the band
$b=\pm 6.4\deg$ along the galactic plane. For details see Winkler
(2001). The present data are from 1199 pointings with an average
exposure of 1400 seconds per pointing.

As a result of data sharing agreements, the results presented here are
limited to the galactic quadrant $l=270\deg$ -- $360\deg$ but, in accordance
with those agreements, data from pointings in the entire GCDE region
$l=\pm 30\deg$ have been taken into account in the analysis.

For each pointing and each of the 19 Ge detectors, the SPI event data
were gain corrected and binned into 0.25 keV wide bins.
The gain correction was performed for each orbit to account for long-term
gain drifts that arise from temperature variations of the detectors.
The instrumental background lines employed for the calibration were 
those at
438.64 keV ($^{69}$Zn),
1107.01 keV ($^{69}$Ge), and
1778.97 keV ($^{28}$Al).
Within this range a linear relationship between raw channel and calibrated 
energy has been
found sufficient and leads to an absolute energy calibration of better than
0.05 keV at 511 keV. Note that the data were taken in a period
immediately following the first SPI detector annealing procedure, when
the energy resolution was optimal.
On average, we obtained an instrumental energy resolution of 
2.16~$\pm$~0.03~keV
full width at half maximum (FWHM) at 511 keV.

The instrumental background at the energy of interest is comprised
of a strong instrumental 511 keV line superimposed on a
continuum spectrum. Typically, the expected
signal-to-background ratio for the observation of the GC
is 2.5\%  for the line only, and of the order of 1.25\%
when also taking into account the underlying continuum.

The instrumental background in the 511 keV line region has been
modelled by two components. The first model component
describes the continuum below the instrumental 511 keV line and is
based on the background level in energy bands adjacent to the
line. The second model component describes the instrumental 511
keV line in intensity and shape and is based on observations of an
empty field that is assumed to be free of celestial 511 keV line
emission. The time variation of both components has been modelled
using the rate of saturating events in the Ge detectors.
Detailed studies of the time variability of the instrumental SPI
background have shown that over wide energy bands, and in
particular for the instrumental 511 keV line, the background
variations follow the variations of the saturating event rate.
This correlation can be understood as follows. Impinging
cosmic-ray protons produce high energy secondary particles (p, n,
$\pi^+$, $\gamma$...) in inelastic interactions with instrument
nuclei. When the primary or secondary particles release more than
$\sim$8~MeV in a Ge detector they trigger a saturating event.
The positrons responsible for the 511 keV background line are
related to this secondary flux since they are produced (1) by the
decay of $\beta^+$ isotopes induced by nuclear reactions of
secondary n and p with instrument material, (2) by pair creation
from secondary $\gamma$'s and (3) by the decay of $\pi^+$.
Similarly, the continuum background rate underneath the line
varies with the saturating event rate since it is mainly due to
the decay of radioactive isotopes produced in Ge detectors. The
yield of these isotopes is mostly proportional to the secondary
neutron flux impinging on Ge detectors (see Naya et~al.
1996).

The continuum  component was taken to be independent of energy but
to follow the number, $S_{p,d}$, of saturating events in the
detector. Thus the predicted number of continuum background
counts, $B_{p,d,e}^{\rm cont}$,  in detector $d$ and energy bin
$e$ during pointing $p$ is given by
\begin{equation}
  B_{p,d,e}^{\rm cont} = F \Delta_e
                          \frac{S_{p,d}}
                               {\sum_{p'} S_{p',d}}
                          \sum_{p'}
                          \frac{\sum_{e'} \displaystyle N_{p',d,e'}}
                               {\sum_{e'} \displaystyle
                               \Delta_{e'}}.
\end{equation}
Here $\Delta_e$ is the width of the spectral bin (in keV) , and
$N_{p,d,e}$ is the number of observed counts. The sum over $e'$ is
taken over the adjacent energy intervals 485-500~keV and
520-550~keV and that over $p'$ over all GC region pointings.
The normalising factors in the denominators have been defined such
that the factor $F$ is nominally unity.

  The level and spectral shape of the instrumental 511 keV line
component was determined from observations of empty fields,
labelled ``off", and is given by
\begin{equation}
  B_{p,d,e}^{\rm line} = G\Delta_e
                     \frac{S_{p,d}}
                          {\sum_{p'} S_{p',d}^{\rm off}}
                     \sum_{p'} \left(
                     \frac{\displaystyle N_{p',d,e}^{\rm off}}
                          {\displaystyle \Delta_{e}} -
                     \frac{\sum_{e'} \displaystyle N_{p',d,e'}^{\rm off}}
                          {\sum_{e'} \displaystyle \Delta_{e'}} \right)
\end{equation}
where $S_{p,d}^{\rm off}$ and $N_{p,d,e}^{\rm off}$ are the number of saturating 
events, and the number of observed counts for the empty field
observation. The sum over $e'$ is taken over the same adjacent
energy intervals that were used for the continuum component while
$p'$ now runs over all ``off'' pointings. The factor $G$ is
analogous to the $F$ in equation~1.

\begin{figure}[tb]
   \includegraphics[width=8.8cm, height=6.0cm]{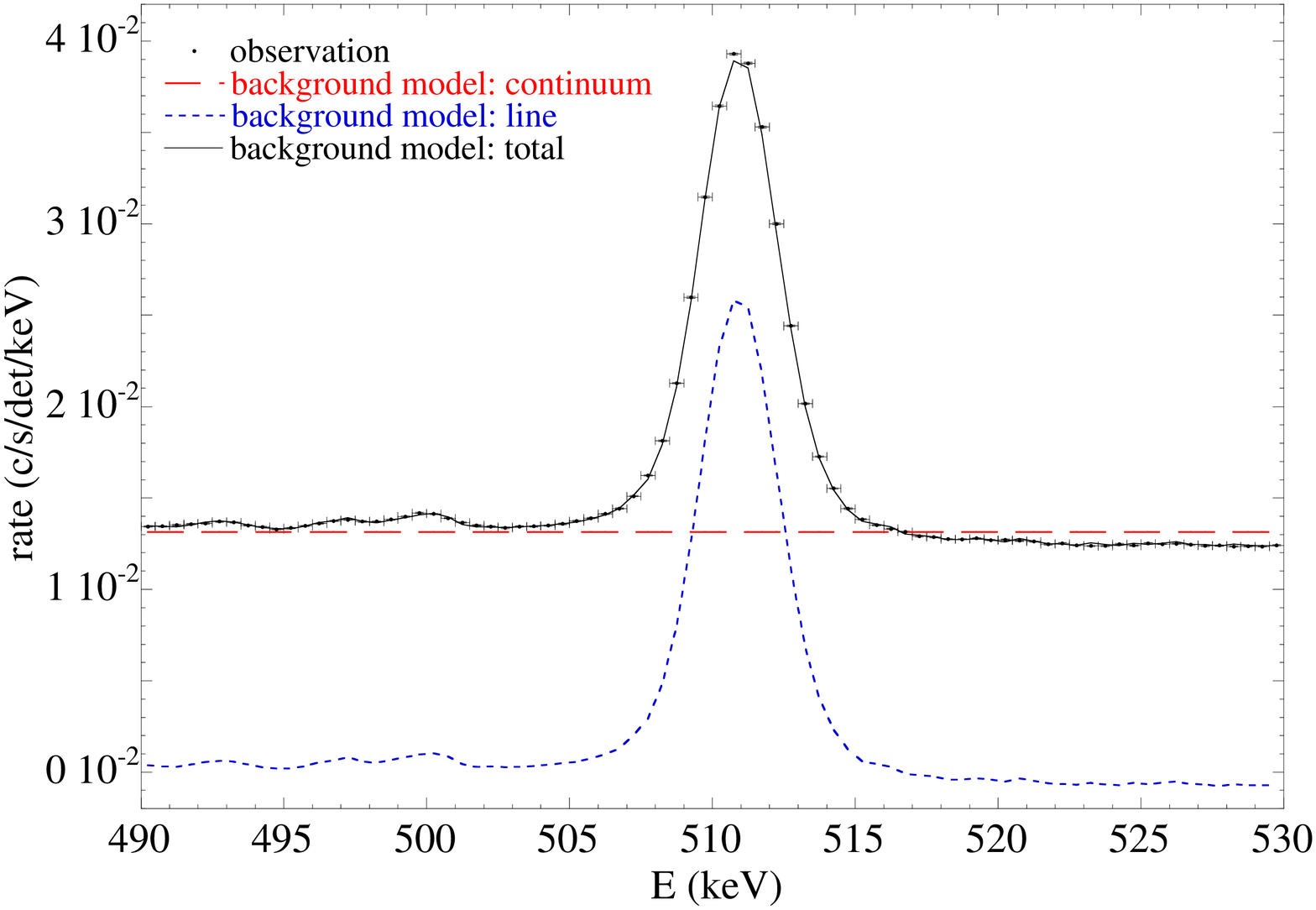}
   \caption{Raw spectrum and background model components.}
    \label{fig:rawspectra}
\end{figure}

Figure \ref{fig:rawspectra} shows the raw spectrum (summed over
all pointings and detectors) together with the two background
model components.

To assess the systematic uncertainties of our approach, we used 3 different
empty field observations for background modelling:
(a) a 820 ks observation of the Cygnus region performed during the
performance and verification phase,
(b) a 718 ks observation of the LMC region, and
(c) a 450 ks observation of the Crab nebula performed after the first
annealing.

The use of the rate of saturated event rates was found to predict
short term variations very well, but if the factors $F$ and $G$
are held at unity, systematic residuals with an amplitude of up to
2\%\ are found over times greater than a few days. This leads to
relatively large uncertainties in the measurement of the flux of
the cosmic 511~keV line and, since the main morphological
information is contained in the temporal modulation, it could
introduce a systematic error in the determination of its spatial
distribution.

The systematic uncertainties are considerably reduced by
introducing a moderate number of free parameters in the analysis.
It turned out that the morphology study is best performed allowing
a different value of $G$ for each of the 19 orbits. In this way,
only the short-term time variability ($<3$ days) is predicted by
the background model while the long-term variability is regarded
as an unknown. For line profile studies, it is the energy
dependence which is more important and the factors $F$ and $G$
were both fitted on a ``per energy bin" basis. With this background
modelling, the spatial distribution models discussed below all
have acceptable $\chi^2$ values (probability $\ga$0.5).

\section{Results}
\label{sec:results}

\begin{figure}[tb]
   \includegraphics[width=8.8cm, height=5.8cm]{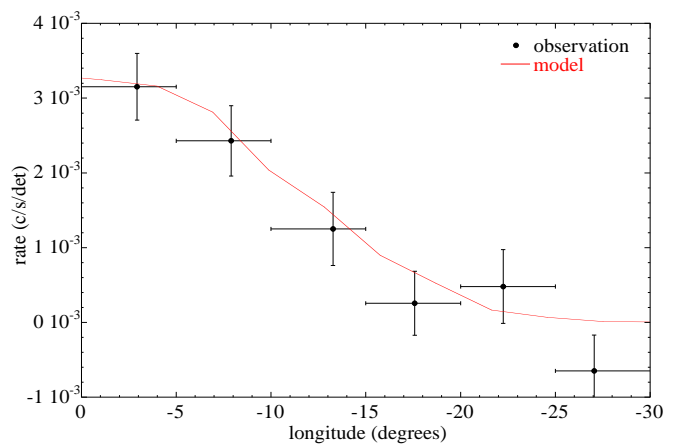}
   \caption{Rate induced by galactic 511~keV photons as a function of 
longitude.
    The response to a gaussian source (FWHM = 10\deg) is also shown for 
comparison.}
    \label{fig:rawrate}
\end{figure}

A first indication of the distribution of the emission can be
obtained by plotting the mean background subtracted counting rate
in the line, averaged over all detectors, as a function of
galactic longitude. This is shown in Fig. \ref{fig:rawrate}. The
profile expected given the 16\deg\ FWHM field of view, for gaussian
shaped source of 10\deg\ FWHM is also shown\footnote{In accordance with
the above-mentioned data-sharing agreement, only negative longitudes are 
shown.}.

A more quantitative approach is possible by taking into account
the spatial and temporal coding by the instrument and fitting to
the data a model of the spatial distribution that has one or more
components. The components can be point sources, gaussian or other
geometric forms, or arbitrary maps corresponding to known
distributions. The intensities of the components are adjusted to
maximize the likelihood that the model gives rise to the observed
distribution of counts in the line (a band 508-514 keV was used),
binned by detector and by pointing. A background, computed as
described in section \ref{sec:analysis}, is included in the
modelling.

Models based on gaussian functions centred on the GC
require a FWHM between $6\deg$ and $18\deg$ ($2\sigma$ limits), with a
best fit FWHM of $10\deg$ (the quoted range includes systematic
uncertainties introduced by the background modelling). Models
based on uniform emission within a given radius, centred again on
the GC, require a diameter between $10\deg$ and
$26\deg$ ($2\sigma$ limits), with an optimum diameter of $14\deg$.

Analyses with models consisting only of point sources, at positions
found in successive iterations by the SPIROS software (Skinner \& 
Connell 2003),
show that a single point source is inconsistent with the data.
Formally, we cannot exclude the possibility that the emission originates in 
at least 2 point sources.

\begin{figure}[tb]
   \includegraphics[width=8.8cm,height=5.0cm]{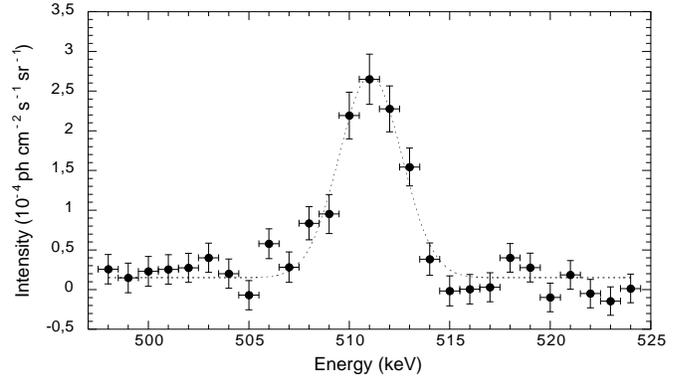}
   \caption{511 keV flux spectrum obtained using a gaussian
   centred on the GC with a FWHM of 10\deg.}
    \label{fig:spectrum}
\end{figure}

To study the 511 keV line profile the data in 1~keV wide energy
bins were independently fitted using models of celestial 511~keV
gamma-ray line emission distributions on top of the background
model. As an example, the resulting photon spectrum obtained using
a 10\deg\ FWHM gaussian for the spatial distribution (our best
fitting model distribution) is shown in Fig.~\ref{fig:spectrum}.
Line profile parameters have been extracted from these spectra by
fitting to the flux values a gaussian function on top of a
constant. Addition of a positronium step does not significantly
change the result.

Using gaussian models with FWHM that vary between our $2\sigma$
limits of 6\deg\ -- 18\deg, we obtain a 511 keV line flux of
$9.9^{+4.7}_{-2.1} \times 10^{-4}$ ph cm$^{-2}$ s$^{-1}$,
a line centroid of $511.06^{+0.17}_{-0.19}$ keV and an observed
line width of $3.66^{+0.36}_{-0.41}$ keV (FWHM).
We note that, in particular for the flux estimates, the uncertainties 
quoted are dominated by the range of models considered. For a specific 
model they are very much smaller (see below).
Subtracting quadratically the instrumental line width of
$2.16 \pm 0.03$ keV at 511~keV results in an intrinsic width of
$2.95^{+0.45}_{-0.51}$ keV (FWHM) for the astrophysical line.
Using different methods of treating the instrumental background
(different ``off" observations and different adjacent energy bands)
provides us with an estimate for the systematic uncertainties in these
measures.
They amount to $\sim10^{-5}$ ph cm$^{-2}$ s$^{-1}$ for the line flux,
$0.03$ keV for the line centroid, and $0.1$ keV for the line width.

Trying alternative celestial intensity distribution models mainly affects
the 511~keV line flux, while the line position and line width always stay
within the quoted uncertainties.
For example, if one fits a point source at the GC -- despite
the fact that such a model is excluded by the data -- this results
in a flux of $(5.4 \pm 0.5) \times 10^{-4}$  ph cm$^{-2}$ s$^{-1}$.
Fitting a uniform emission with a diameter of $14\deg$ around the GC
-- a model that fits the data almost as well as the $10\deg$
gaussian -- results in a flux of $(9.4 \pm 0.8) \times 10^{-4}$
ph cm$^{-2}$ s$^{-1}$.

\section{Conclusions}
\label{sec:conclusions}

Preliminary results from the early mission phase show that SPI on
INTEGRAL allows a study of the GC 511~keV emission with an unprecedented combination of spectral
and angular resolution and of sensitivity. In general the
results are consistent with earlier measurements (see Table~1). We can 
exclude with high significance
models in which the source is relatively compact; for a gaussian source
centred at $l=0\deg$, $b=0\deg$ the 2$\sigma$ lower limit on the FWHM is 6\deg.
This is in the high side of the 3.9\deg\ -- 5.7\deg\ FWHM range derived
from OSSE, SMM and TGRS data (Milne et~al. 2000).
In the spectral domain, our measurements show that the line width
(2.95~$^{+0.45}_{-0.51}$~keV) is at the upper bound of the range of values 
previously
reported.

In this short communication we have confined ourselves to simple
spatial and spectral descriptions of the data. Constraints on
distributions suggested by earlier measurements with poorer
spatial and/or spectral sensitivity and studies of the positronium
emission (evidence for which can perhaps be seen in Fig. \ref{fig:spectrum}) will be 
discussed elsewhere.

\begin{acknowledgements}
The SPI project has been completed under the responsability and
leadership of CNES. We are grateful to CEA, CNES, DLR, INTA and NASA
for support.
\end{acknowledgements}


\end{document}